# The tight binding approximation and thermodynamic functions

B.V. Karpenko[a], A.V. Kuznetzov[b] and V.V. Dyakin[c]

a) Karpenko Boris Victorovich, Institute of Physics of Metals, URO RAN, Ekaterinburg 620041, Russia, phone: (343) 378-35-64, fax: (343) 374-52-44, e-mail: boris.karpenko@mail.ru
b) Kuznetzov Alexander Vasilyevich, Ural State University, Ekaterinburg 620083, Russia, phone: (343) 218-05-77, e-mail: al.vas.kuz@gmail.com
c) Dyakin Viliam Vyacheslavovich, Institute of Physics of Metals, URO RAN, Ekaterinburg 620041, Russia, phone: (343) 370-05-44, fax: (343) 374-52-44

**Abstract**

The various thermodynamic functions dependence on degree of energy band occupation and temperature was studied. The one-band tight binding approximation for the electron energy spectrum was used. The Fermi energy, density of states, degeneracy temperature, chemical potential, partition function, thermodynamic potential, energy, free energy, entropy, heat capacity, spin magnetization and initial susceptibility were calculated. The limited energetic spectrum leads to the peculiarities in the behavior of these quantities in comparison with free electron gas.

**1. Introduction**
Considering the thermodynamics of the delocalized electron system in solids one must describe energy spectrum in the terms of band theory and use the Fermi statistics. The tight binding approximation is one of the approaches. The numerical calculations of the above mentioned thermodynamic functions have practical and academic interest.

This paper is an extended version of previous article [1].

**2. Spectrum, concentration, distribution function**
It is known that electron energy $E$ in tight binding nearest neighbor approximation in the simple cubic lattice has the form

$$E = \Delta \cdot \varepsilon, \qquad (1)$$

where $\Delta$ is the bandwidth,

$$\varepsilon = \frac{1}{6}(3 - \cos x - \cos y - \cos z), \quad -\pi \leq x, y, z \leq \pi. \qquad (2)$$

The variables $x, y, z$ are the components of dimensionless quasi-momentum. The model under consideration belongs to the limited spectrum problems.
Below we will study the system with the arbitrary degree of energy band filling that is with electron concentration $n$ at

$$0 \leq n \leq 2 \qquad (3)$$

per one lattice site on the account of the spin. We write Fermi distribution function using the dimensionless variables:

$$f = \frac{1}{1 + \exp\frac{1}{t}(\varepsilon - w)} \quad , \quad t = \frac{kT}{\Delta} \quad , \quad w = \frac{\mu}{\Delta} \, , \qquad (4)$$

where $\mu$ is the chemical potential and $\varepsilon$ is given in Eq. (2). The relation between $w$ and $n$ is determined by the usual equation

$$\frac{2}{(2\pi)^3} \int_{-\pi}^{\pi} \int_{-\pi}^{\pi} \int_{-\pi}^{\pi} f dx dy dz = n. \qquad (5)$$

### 3. Fermi energy, density of states

Fermi energy $\varepsilon_F$ is determined by the equation

$$n = \frac{2}{(2\pi)^3} \iiint dx dy dz. \qquad (6)$$

The volume of integration in Eq. (6) is limited by the isoenergetic surface $\varepsilon_F$. The function $\varepsilon_F(n)$ is shown in Fig. 1.

We use the formula

$$D(\varepsilon) = \frac{2}{8\pi^3} \int \frac{dS}{|\nabla \varepsilon|} \qquad (7)$$

for density of states. In Eq. (7) the integration is over the isoenergetic surface $\varepsilon$. (The article [1] has a misprint in Eq. (7), it lacks the factor of "2" on the right side) The well known graph of $D(\varepsilon)$ is represented in Fig. 2.

### 4. Thermodynamic functions

Degeneracy temperature $t_g$ is determined from the condition of equality to zero for chemical potential $w$. The Fig. 3 shows the dependence degeneracy temperature on concentration. The magnitude $t_g(n)$ diverges:

$$t_g = \frac{1}{4(1-n) + \frac{[4(1-n)]^3}{32} + ..}. \qquad (8)$$

At $n \to 0$ we have

$$t_g \approx \left\{ \frac{\pi^{3/2}\sqrt{2}(1+\sqrt{2})}{6\sqrt{3}\varsigma(3/2)} n \right\}^{2/3} = 0.78857927 n^{2/3} \qquad (9)$$

and at $n \to 2$ the asymptotic is:

$$t_g \approx -\left\{ \frac{\pi^{3/2}\sqrt{2}(1+\sqrt{2})}{6\sqrt{3}\varsigma(3/2)} (2-n) \right\}^{2/3} = -0.78857927(2-n)^{2/3} \qquad (10)$$

($\varsigma(3/2)$ is Riemann $\varsigma$-function).

The chemical potential $w$ was determined from Eq. (5). The family of curves $w(t,n)$ is represented in Fig. 4. Chemical potential diverges at $n=2$. At high temperatures the chemical potential is given by the asymptotical formula

$$w = \frac{1}{2} + t\ln\frac{n}{2-n} + \frac{n-1}{48}\frac{1}{t} + (n-1)O\left(\frac{1}{t^3}\right) \qquad (11)$$

and the distribution function $f$ is approximated by the expression

$$f = \frac{n}{2}\left\{1 - \frac{(2-n)(2\varepsilon-1)}{4}\frac{1}{t} + \frac{(2n-2)(1-n)}{4}\left[\frac{5}{24} - \varepsilon + \varepsilon^2\right]\frac{1}{t^2}\right\}. \qquad (12)$$

At zero temperature
$$w = \varepsilon_F \qquad (13)$$
and $f$ equals unity at $\varepsilon \leq \varepsilon_F$ and zero at $\varepsilon \geq \varepsilon_F$ (usual step function).

Thermodynamic potential $\Omega$ was found from the expression

$$\Omega = -2\frac{t}{(2\pi)^3}\int_{-\pi}^{\pi}\int_{-\pi}^{\pi}\int_{-\pi}^{\pi}\ln\left(1+\exp\frac{1}{t}(w-\varepsilon)\right)dxdydz. \qquad (14)$$

The curves $\Omega(t,n)$ are shown in Fig. 5. The magnitude $\Omega$ diverges at $n=2$. At the high temperatures

$$\Omega = -2t\left\{\ln\frac{2}{2-n} + \frac{1}{192t^2}\frac{n-(n-1)^2(2-n)}{(2-n)}\right\}. \qquad (15)$$

At zero temperature the potentials $\Omega(n)$ was calculated by the relation
$$\Omega(n) = E_g(n) - n\varepsilon_F(n), \qquad (16)$$
where $E_g$ is the ground state energy (see below).

We have for partition function $Z$ the formula
$$\ln Z = -\frac{1}{t}\Omega. \qquad (17)$$

The corresponding graphs are in Fig. 6. The magnitude diverges for $n=2$.
The asymptotic formulas are obtained by dividing Eqs. 15 and 16 with (-t).
At zero temperature the magnitudes $\ln Z(n)$ diverge.

The energy $E$ was found from the equation

$$E = \frac{2}{(2\pi)^3}\int_{-\pi}^{\pi}\int_{-\pi}^{\pi}\int_{-\pi}^{\pi}\varepsilon f\,dxdydz. \qquad (18)$$

Fig. 7 demonstrates the graphs of $E(n,t)$. At the high temperatures

$$E = \frac{n}{2}\{1 - \frac{1}{24}(2-n)\frac{1}{t} + O(\frac{1}{t^3})\}. \qquad (19)$$

At zero temperature the energy (ground state energy $E_g$) was calculated by Eq. (18), using the step function for $f$ and the expression Eq. (7) for density of states.

The free energy $F$ is determined as
$$F = \Omega + wn. \qquad (20)$$

For convenience the graphs are represented separately for the cases $n \leq 1$ and $n \geq 1$ in Figs. 8 and 9. The high temperature expression is

$$F = t\left\{n \ln \frac{n}{2} + (2-n) \ln \frac{2-n}{n}\right\} + \frac{n}{2} - \frac{1-(n-1)^2}{96t}. \tag{21}$$

At zero temperature the free energies $F(n)$ equal to the corresponding energies $E(n)$.

The entropy $S$ we determine from the formula

$$S = \frac{1}{t}(E - F). \tag{22}$$

The curves of $S(n,t)$ are given in Fig. 10 for the concentrations mutually complementary for 2. In the high temperature limit

$$S = (2-n) \ln \frac{2}{2-n} + n \ln \frac{2}{n} + \frac{4-5n}{6t} - \frac{(2-n)n}{32t^2}. \tag{23}$$

At zero temperature the entropy equals zero.

The heat capacity is given by the expression

$$C = \frac{dE}{dt}. \tag{24}$$

The graphs of $C(n,t)$ are represented in Fig, 11 for the concentration mutually complementary for 2. At the high temperatures

$$C = \frac{n}{4t}\{(2-n)\ln\frac{2-n}{n} + \frac{(2-n)(n-1)}{12}\ln\frac{2-n}{n}\frac{1}{t} +$$

$$+ [\frac{(n-1)(1-2((n-1)^2)}{48n} + \frac{2(n-1)^3 - n(4-3n)}{96n}\ln\frac{2-n}{n}]\frac{1}{t^2}\}. \tag{25}$$

At zero temperature the heat capacity equals zero.

Let us define the magnetization in external magnetic field $H$ as the difference of concentration of electrons for spins directed along the field $n_1$ and against $n_2$:

$$m = n_1 - n_2, \tag{26}$$

$$n_{1,2} = \frac{1}{(2\pi)^3} \int_{-\pi}^{\pi}\int_{-\pi}^{\pi}\int_{-\pi}^{\pi} f_{1,2} dxdydz, \tag{27}$$

$$f_{1,2} = \frac{1}{1+\exp\frac{1}{t}(\varepsilon_{1,2} - w)}, \quad \varepsilon_1 = \varepsilon - h, \quad \varepsilon_2 = \varepsilon + h, \quad h = \mu_B \frac{H}{\Delta}, \tag{28}$$

$$n_1 + n_2 = n. \tag{29}$$

The curves $m(n,h)$ at zero temperature are shown in Fig. 12. The nominal magnetization value $n$ reaches at magnetic fields $h_c$ which are determined by the expressions: $h_c(n) = \frac{1}{2}\varepsilon_F(2n)$ if $n \leq 1$, and $h_c(n) = \varepsilon_F(2(2-n))$ if $n \geq 1$. The dependence $h_c(n)$ is shown in Fig. 13. The curves $m(h)$ become smoother at elevated temperatures.

The initial magnetic susceptibility $\chi_0$ is defined as

$$\chi_0 = \frac{dm}{dh} \text{ at } h=0. \tag{30}$$

The calculations give

$$\chi_0 = \frac{K}{t}, \tag{31}$$

$$K = \frac{2}{(2\pi)^3} \int_{-\pi}^{\pi}\int_{-\pi}^{\pi}\int_{-\pi}^{\pi} \frac{\exp\left(\frac{1}{t}(\varepsilon-w)\right)}{\left(1+\exp\left(\frac{1}{t}(\varepsilon-w)\right)\right)^2} dxdydz. \tag{32}$$

We used condition

$$\frac{dw}{dh} = 0 \text{ at } h=0 \tag{33}$$

when receiving Eqs. (19,20). The chempotential $w$ was given for zero field. The curves $\chi_0(n,t)$ are shown in Fig. 14 and Fig. 15 for two scales. At the high temperatures we have the asymptotic expression

$$\chi_0 = \frac{n(2-n)}{2}\frac{1}{t} + O\left(\frac{1}{t^3}\right). \tag{34}$$

At zero temperature initial susceptibilities equal to the corresponding density of states at the Fermi energy.

We demonstrate the graph of $\chi_0(t)$ for free electron gas for the comparison in Fig. 16.

### 5. Discussion

The thermodynamics of the system with the limited energy spectrum has a number of features when comparing with the system without this limitation. However these peculiarities have the transparent physical sense and are predictable in some cases.

The concentration $n=1$ is the "critical" one in some sense. The degeneracy temperature $t_g$ diverges at this concentration (see Fig. 3) and the chempotential $w$ does not depend on the temperature (see Fig. 4), remaining equal $0.5$ (Fig. 4). In the case of ideal gas the degeneracy temperature makes up $0.9887335$ of Fermi temperature and is proportional to $2/3$ power of volume concentration of electrons. The chempotential may or fall with temperature ($n<1$) or increase ($n>1$). It only decreases in the case of free gas. The Fermi distribution function $f$ at the high temperatures does not take the Boltzmann form and tends to constant value $n/2$ (see Eq. (12)).

The free energy $F$ has different dependence on $t$ for $n<1$ and $n>1$. In the first case we have the complete inversion in arrangement upon $n$, but in the second case the order of curves is conserved with $t$.

The functions $S, C, m, K$ are symmetric relative of the value $n=1$ what demonstrates the equivalence of electrons and holes. The energy $E$ and entropy $S$ tend to saturation with the temperature growth but not rise unlimitedly as in free gas (see Figs. 7 and 10). The specific heat $C$ has the maximum as the function of temperature and tends to zero value as temperature tends to infinity (Fig. 11) in contrast to free gas case, where $C$ is the monotonic function of $t$ which goes to saturation (Dulong and Petit law). The initial susceptibility $\chi_0$ has the low- temperature maximum at some values of $n$ that one may explain by the peculiar density of states dependence of energy in comparison with gas (see Figs. 2, 14, 15, 16). The Pauli susceptibility of electron gas has no maximum at the finite temperatures. At high temperatures our $\chi_0(t)$ asymptotically follows Curie law and the "Curie constant" is proportional to the product of the concentration of electrons $n$ and holes $2-n$ (see Eq. (34)). We want to note that the chemical potential plays the special role in the system thermodynamics since all the other thermodynamic functions depend upon the $w$. Four parameters – band width, concentration, temperature and magnetic field – determine the function under consideration.

**Acknowledgements**
The authors are indebted to Department of Physical Science and Presidium of Ural Branch of the Russian Academy of Sciences, Program "The Physics of new materials and structures", the Russian Foundation for Basic Research (project N 07 02 000 68).

## Appendix

As appears from the air of the used above thermodynamic functions their asymptotic behavior at $t \to \infty$ is determined by the corresponding behavior of the chemical potential $w$. Therefore it is sufficient to show the key moments of deriving the asymptotics of $w$ at $t \to \infty$.

The function $w$ is determined from the equation

$$n = \frac{2}{\pi^3}\int_0^\pi\int_0^\pi\int_0^\pi \frac{dxdydz}{1+\exp\left(\frac{\varepsilon-w}{t}\right)} = \frac{2}{\pi^3}\int_{-\pi/2}^{\pi/2} dx \int_{-\pi/2}^{\pi/2} dy \int_{-\pi/2}^{\pi/2} dz \frac{1}{1+\exp\left(\frac{S(x,y,z)}{6t}+\frac{1/2-w}{t}\right)}, \quad (A.1)$$

where
$$S(x,y,z) = \sin x + \sin y + \sin z. \quad (A.2)$$

We have from Eq. (A.1)

$$1 - n = \frac{1}{\pi^3}\int_{-\pi/2}^{\pi/2}\int\int th\left\{\frac{S(x,y,z)}{12t}+\frac{1/2-w}{2t}\right\}dxdydz. \quad (A.3)$$

Using the equality

$$\int_{-\pi/2}^{\pi/2}\int\int th\frac{S(x,y,z)}{p}dxdydz = 0 \text{ for } 0 < p \le \infty \quad (A.4)$$

we obtain from Eq. (A.3):
$$n = 1 \text{ at } w = 1/2 \quad (A.5)$$
for any $t$.

Introducing the new symbols
$$\sigma = S(x,y,z)/12t, \gamma = \frac{1/2-w}{2t} \quad (A.6)$$

we have Taylor series at $t \to \infty$:

$$th(\gamma+\sigma) = th\gamma + (1-th^2\gamma)\sigma - th\gamma(1-th^2\gamma)\sigma^2 - \frac{1}{3}(1-4th^2\gamma+3th^4\gamma)\sigma^3 +$$

$$+ \frac{1}{3}(2th\gamma - 5th^3\gamma + 3th^5\gamma)\sigma^4 + ... \quad (A.7)$$

With the help of the following equalities

$$\int_{-\pi/2}^{\pi/2}\int\int (\sin x + \sin y + \sin z)^{2k+1} dxdydz = 0, k = 0,1,2..., \quad (A.8)$$

$$\int_{-\pi/2}^{\pi/2}\int\int (\sin x + \sin y + \sin z)^2 dxdydz = \frac{3}{2}\pi^3, \quad (A.9)$$

$$\int_{-\pi/2}^{\pi/2}\int\int (\sin x + \sin y + \sin z)^4 dxdydz = \frac{45}{8}\pi^3 \quad (A.10)$$

we obtain from Eq. (A.3):

$$1-n = x - (x-x^3)\frac{3}{2}\varepsilon + (2x-5x^3+3x^5)\frac{15}{8}\varepsilon^2 + ...; x = th\gamma, \varepsilon = 1/(12t)^2. \qquad (A.11)$$

The solution of the Eq. (A.11) is sought as a $\varepsilon$ degree power expansion:
$$x = x_0 + x_1\varepsilon + x_2\varepsilon^2 + ... \qquad (A.12)$$

From (A.11) and (A.12) one finds:
$$x_0 = 1-n; x_1 = \frac{3}{2}n(1-n)(2-n); x_2 = \frac{3}{8}n(1-n)(2-n)(15n^2 - 12n - 7). \qquad (A.13)$$

Using the definitions for $x$, $\gamma$ and expression
$$arth(z) = \frac{1}{2}\ln\frac{1+z}{1-z} \qquad (A.14)$$

we obtain from Eqs. (A.12) and (A.13)
$$w \approx \frac{1}{2} - t\ln\frac{2-n}{n} - \frac{1-n}{48}\frac{1}{t} + \frac{(n-1)[15n^2 - 12n - 7 + 3(n-1)(n^2 + (2-n)^2)]}{3^3 \times 2^{10}}\frac{1}{t^3}. \qquad (A.15)$$

**References**
[1] Karpenko B.V., Kuznetsov A.V., Dyakin V.V. Energy band and thermodynamics. arXiv:1003.2283v1 [cond-mat.stat-mech]

**Figure captions**
Fig. 1. The dependence of Fermi energy $\varepsilon_F$ on the concentration $n$.

Fig. 2. The graph of density of states $D(\varepsilon)$.

Fig. 3. The degeneracy temperature $t_g$ as a function of concentration $n$.

Fig. 4. The dependencies of chempotential $w$ on temperature at different concentrations $n$. The curves are arrange upwards at the $n = 0.1, 0.2 ... 1.9$ and $1.99$.

Fig. 5. The dependencies of potential $\Omega$ on temperature at different concentrations $n$. The curves are arrange downwards at the $n = 0.1, 0.2 ... 1.9$ and $1.99.$.

Fig. 6. The dependencies of $\ln Z$ on temperature at different concentrations $n$. The curves are arrange upwards at $n = 0.1, 0.2 ... 1.9$ and $n = 1.99$.

Fig. 7. The dependencies of energy $E$ on temperature at different $n$. The curves are arrange downwards at $n = 0.1, 0.2 ... 2$.

Fig. 8. The free energy $F$ as a function of temperature at $n \leq 1$. The concentration increases upwards at $t = 0$.

Fig. 9 . The free energy $F$ as a function of temperature at $n \geq 1$. The curves are arranged upwards at $n = 1, 1.1 ... 2$.

Fig. 10. The entropy $S$ as a function of temperature. The curves are arranged upwards at $n = 0.1(1.9), 0.2(1.8) ... 1$.

Fig. 11. The heat capacity $C$ as a function of temperature. The curves are arranged upwards at $n = 0.1(1.9), 0.2(1.8) ... 1$.

Fig. 12. The dependences $m(h)$. The curves are arranged upwards at $n = 0.1(1.9), 0.2(1.8) ... 1$.

Fig. 13 The dependence $h_c(n)$.

Fig. 14. The dependences $\chi_0(n,t)$ (see text).

Fig. 15. The dependences $\chi_0(n,t)$ at low temperatures (see text).

Fig. 16. The dependence $\chi_0(t)$ for the free electron gas ($t = k_B T / E_F$).

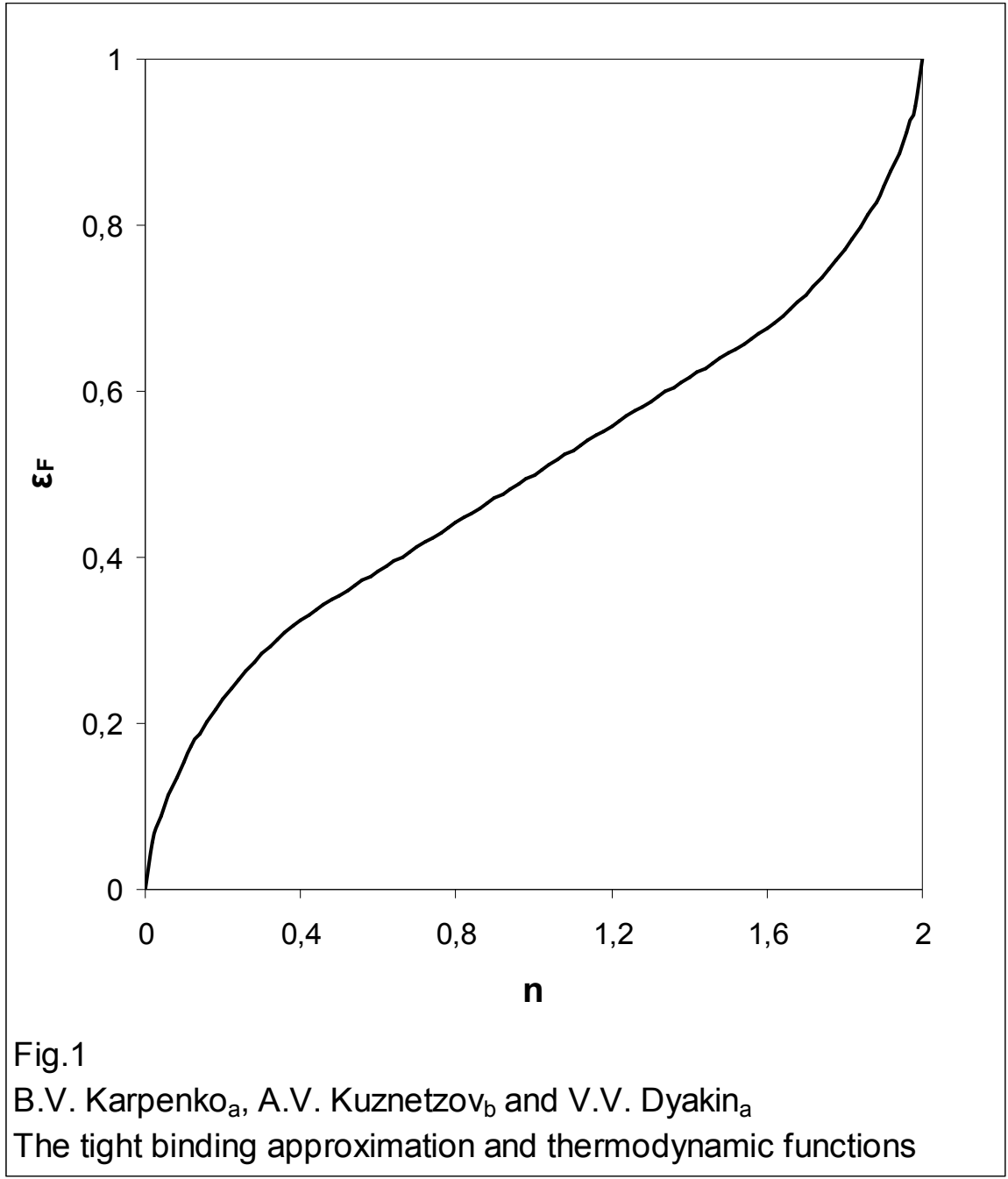

Fig.1
B.V. Karpenko[a], A.V. Kuznetzov[b] and V.V. Dyakin[a]
The tight binding approximation and thermodynamic functions

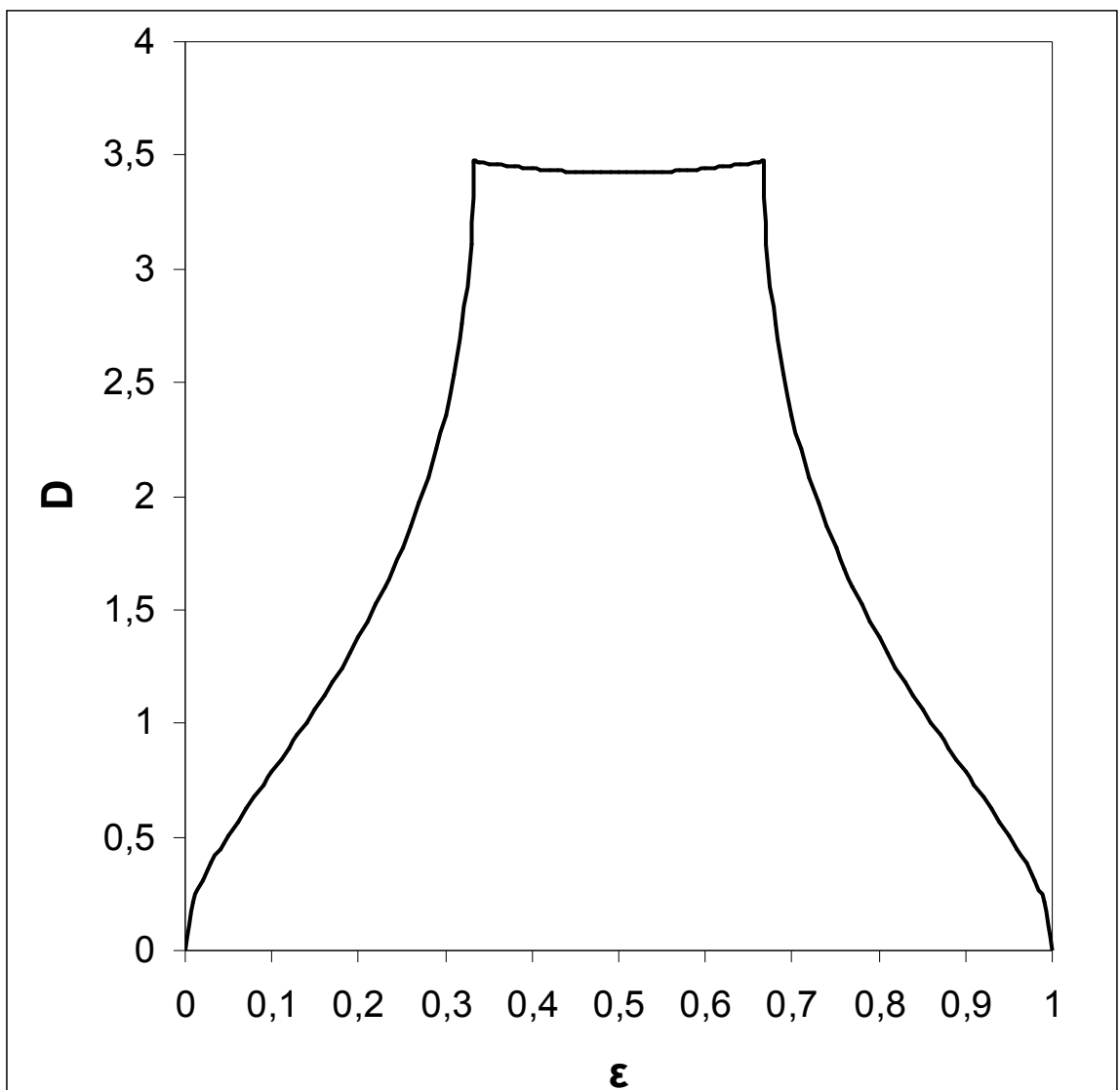

Fig.2
B.V. Karpenko[a], A.V. Kuznetzov[b] and V.V. Dyakin[a]
The tight binding approximation and thermodynamic functions

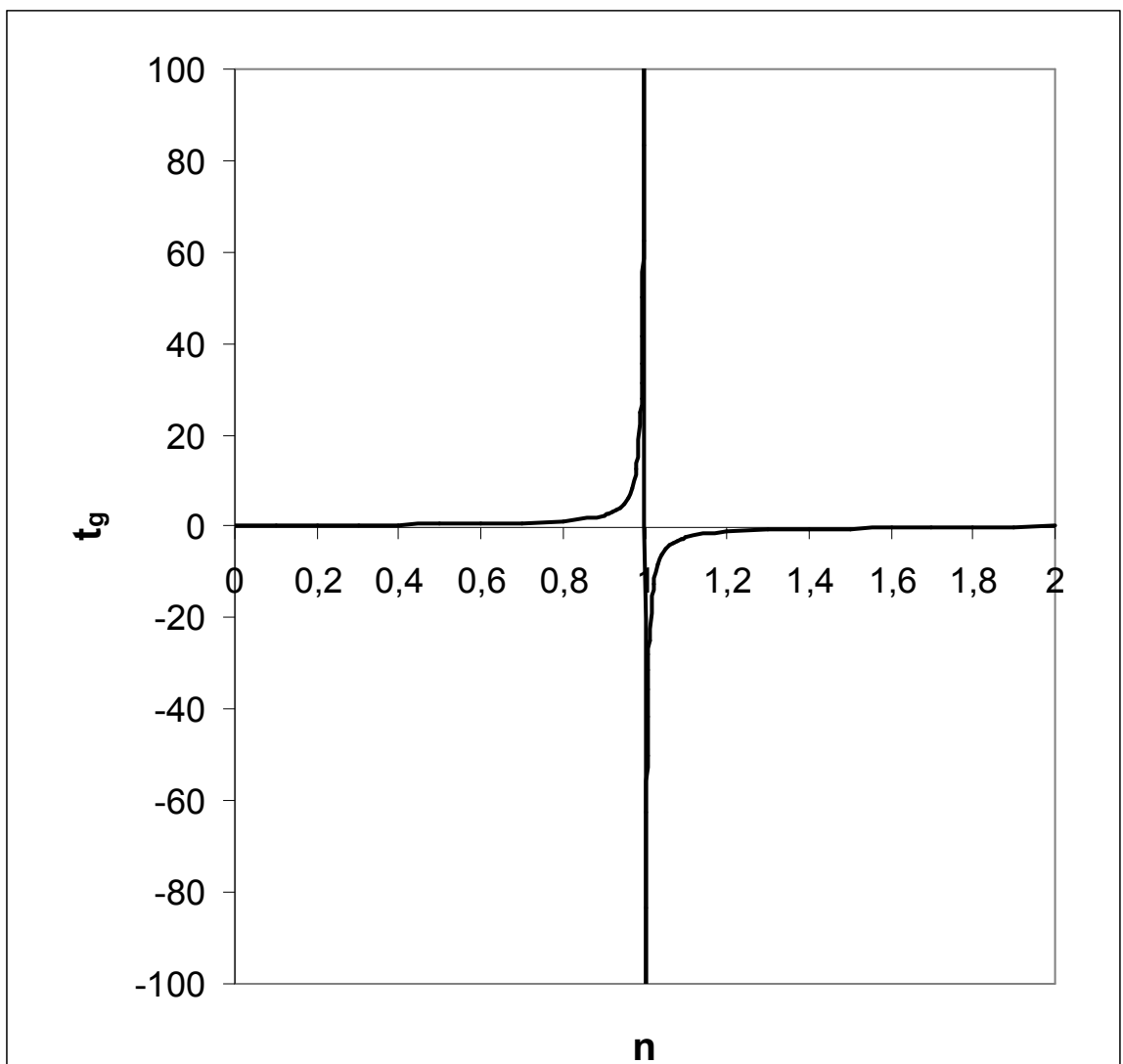

Fig.3
B.V. Karpenko[a], A.V. Kuznetzov[b] and V.V. Dyakin[a]
The tight binding approximation and thermodynamic functions

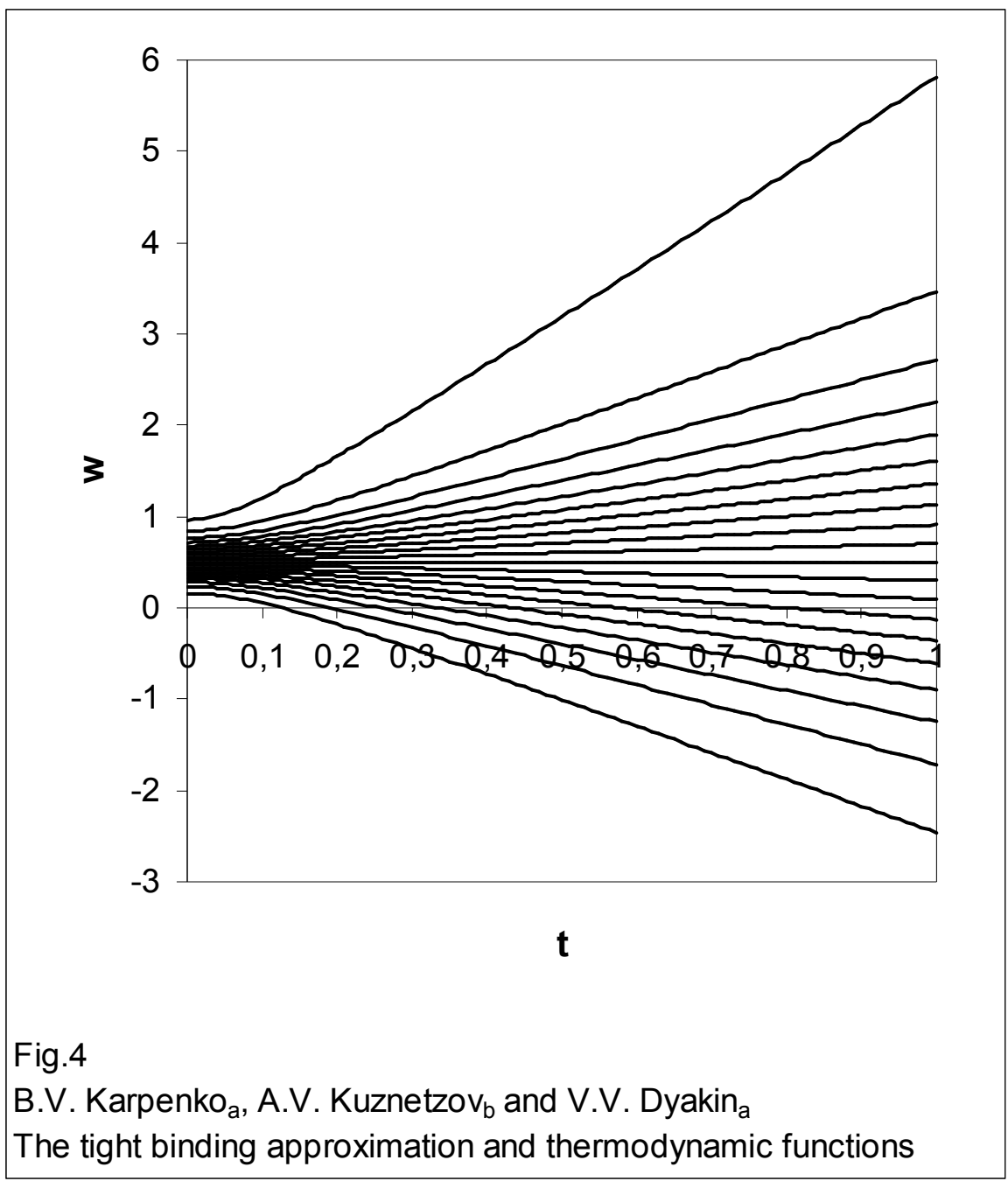

Fig.4
B.V. Karpenko[a], A.V. Kuznetzov[b] and V.V. Dyakin[a]
The tight binding approximation and thermodynamic functions

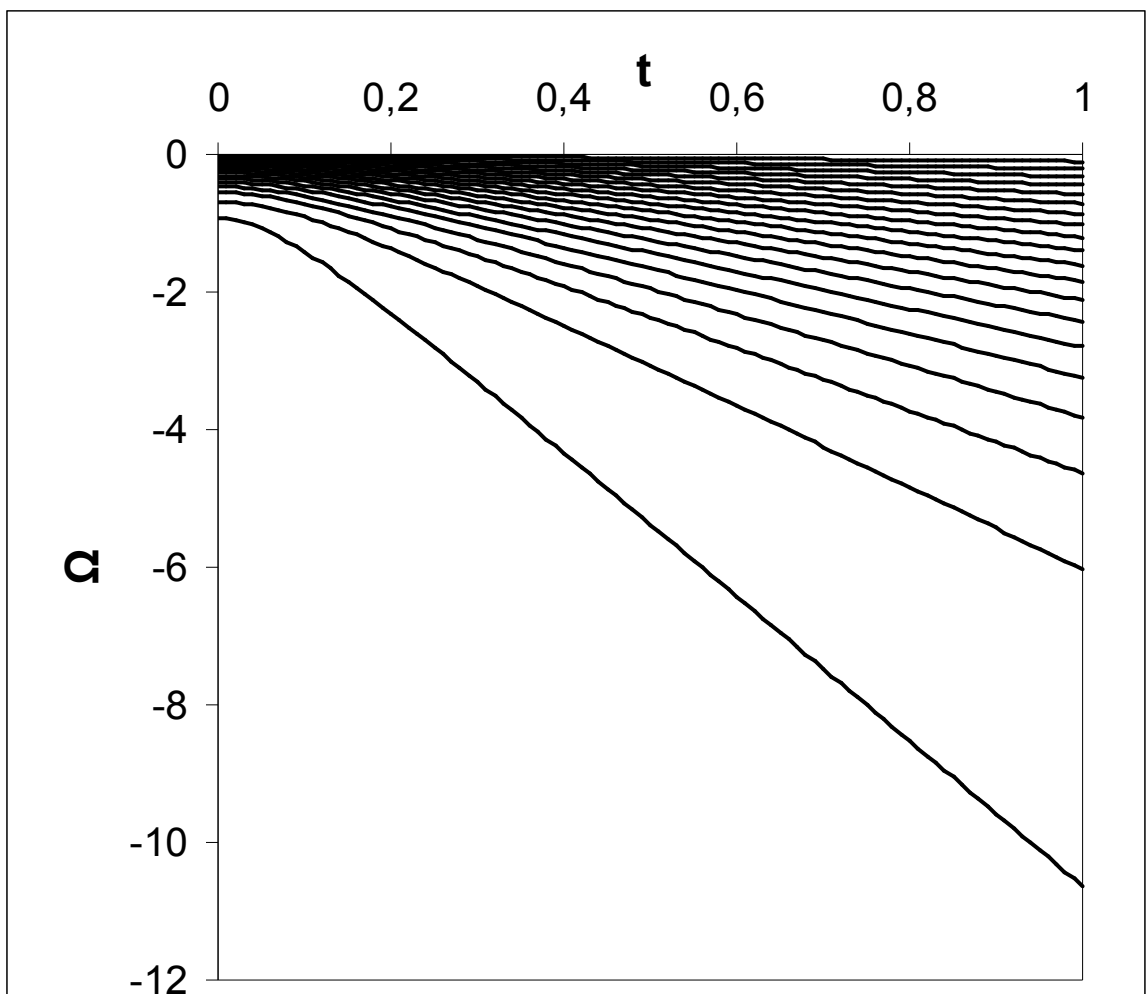

Fig.5
B.V. Karpenko[a], A.V. Kuznetzov[b] and V.V. Dyakin[a]
The tight binding approximation and thermodynamic functions

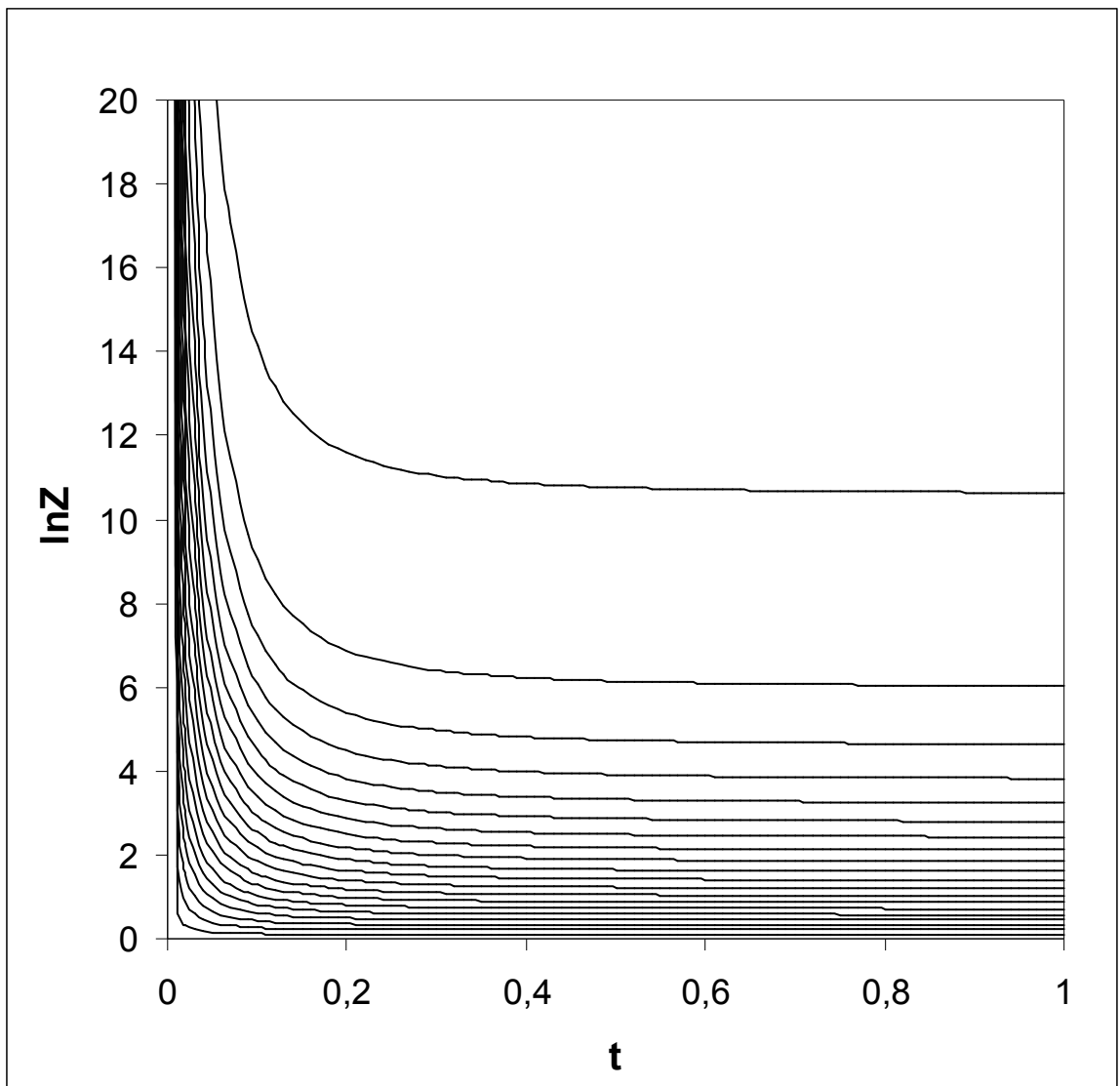

Fig.6
B.V. Karpenko[a], A.V. Kuznetzov[b] and V.V. Dyakin[a]
The tight binding approximation and thermodynamic functions

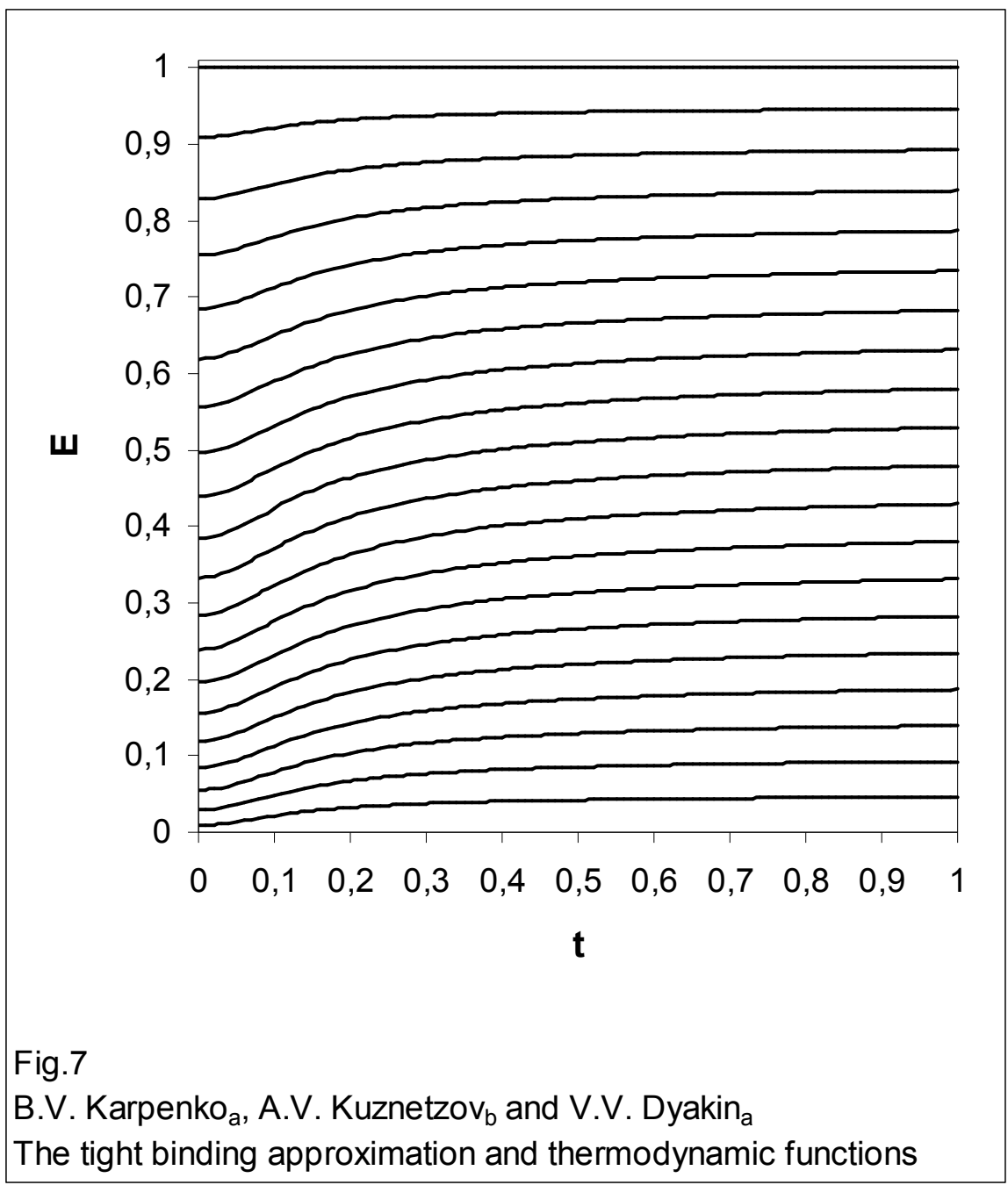

Fig.7
B.V. Karpenko[a], A.V. Kuznetzov[b] and V.V. Dyakin[a]
The tight binding approximation and thermodynamic functions

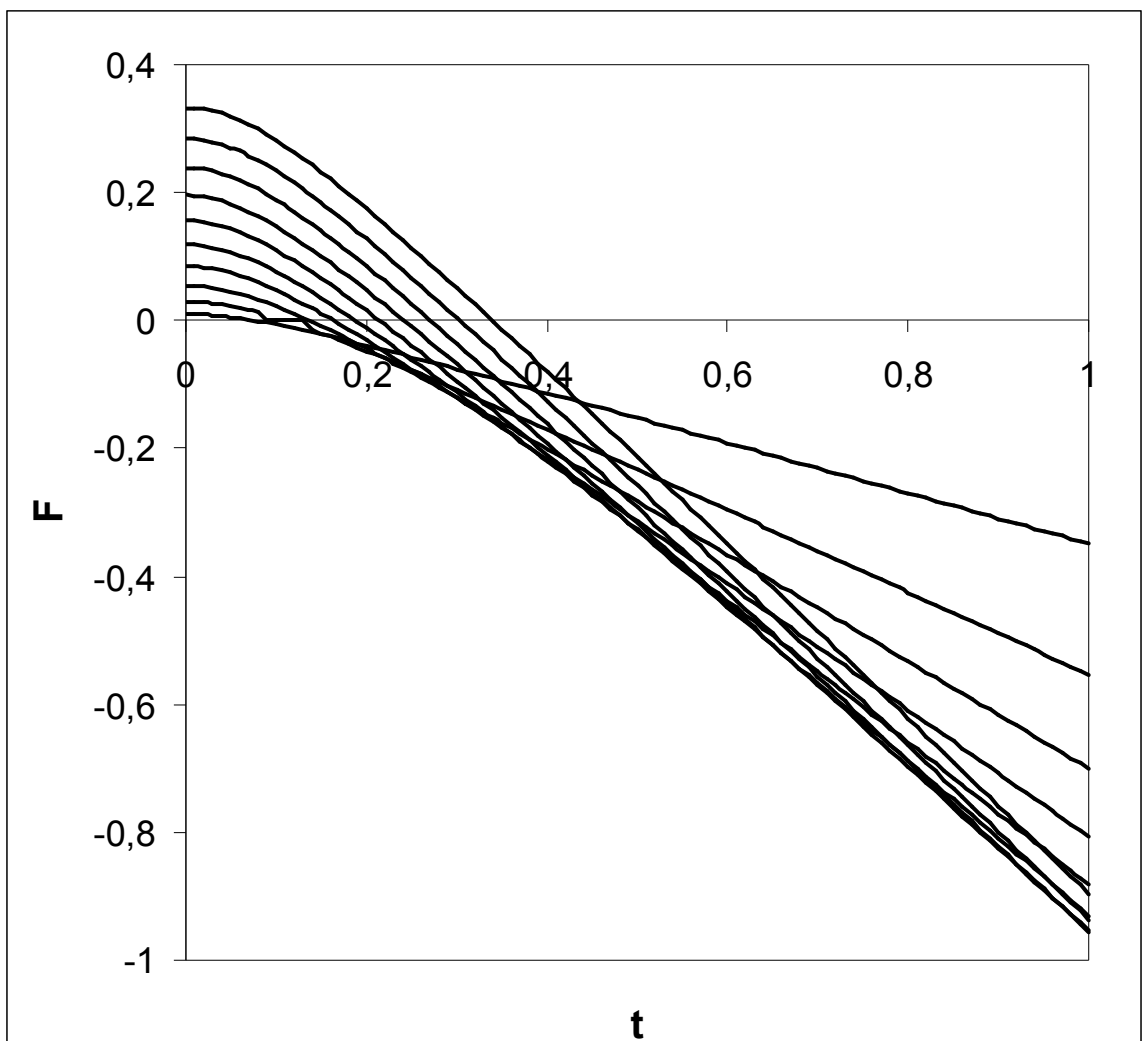

Fig.8
B.V. Karpenko[a], A.V. Kuznetzov[b] and V.V. Dyakin[a]
The tight binding approximation and thermodynamic functions

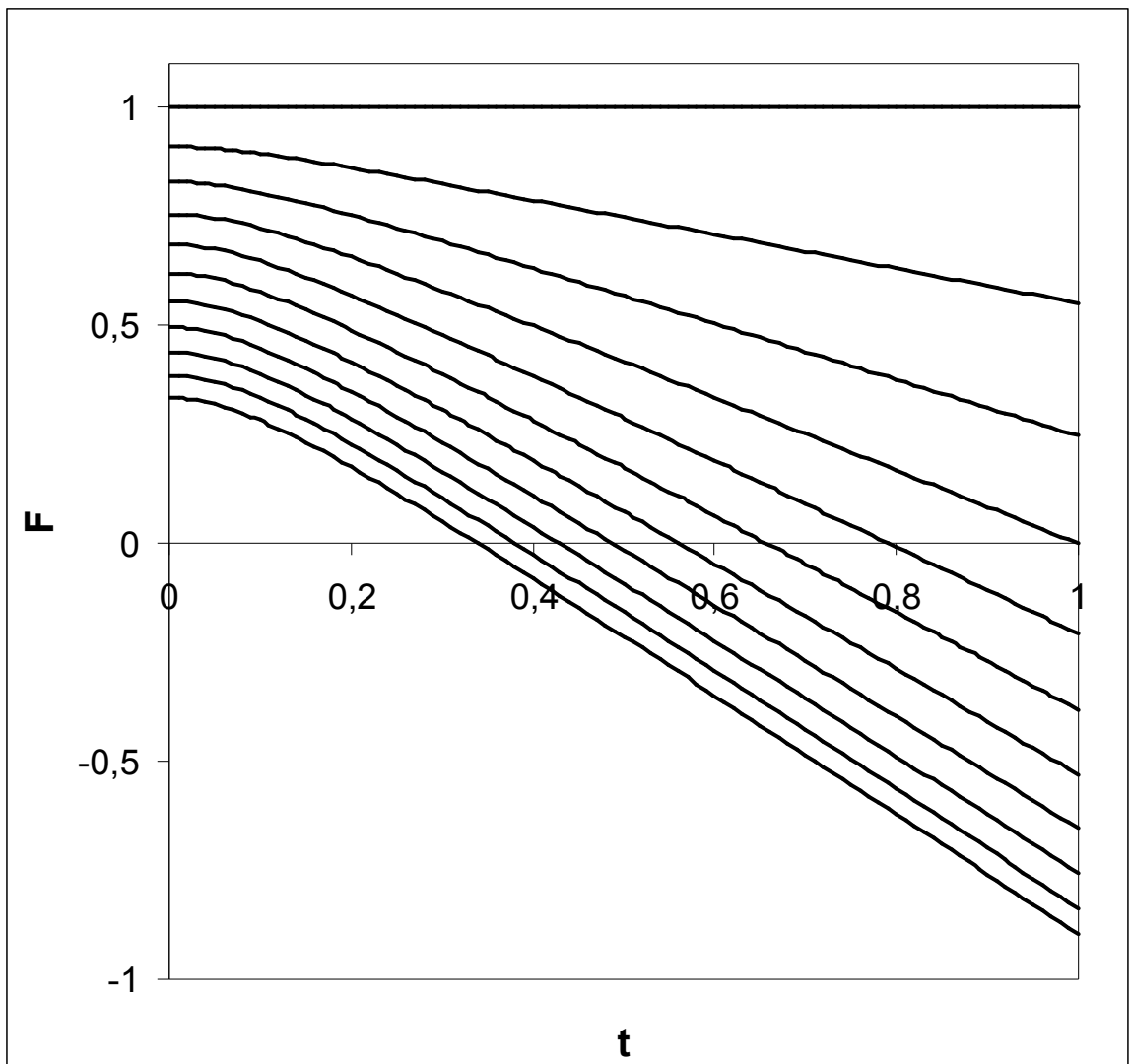

Fig.9
B.V. Karpenko[a], A.V. Kuznetzov[b] and V.V. Dyakin[a]
The tight binding approximation and thermodynamic functions

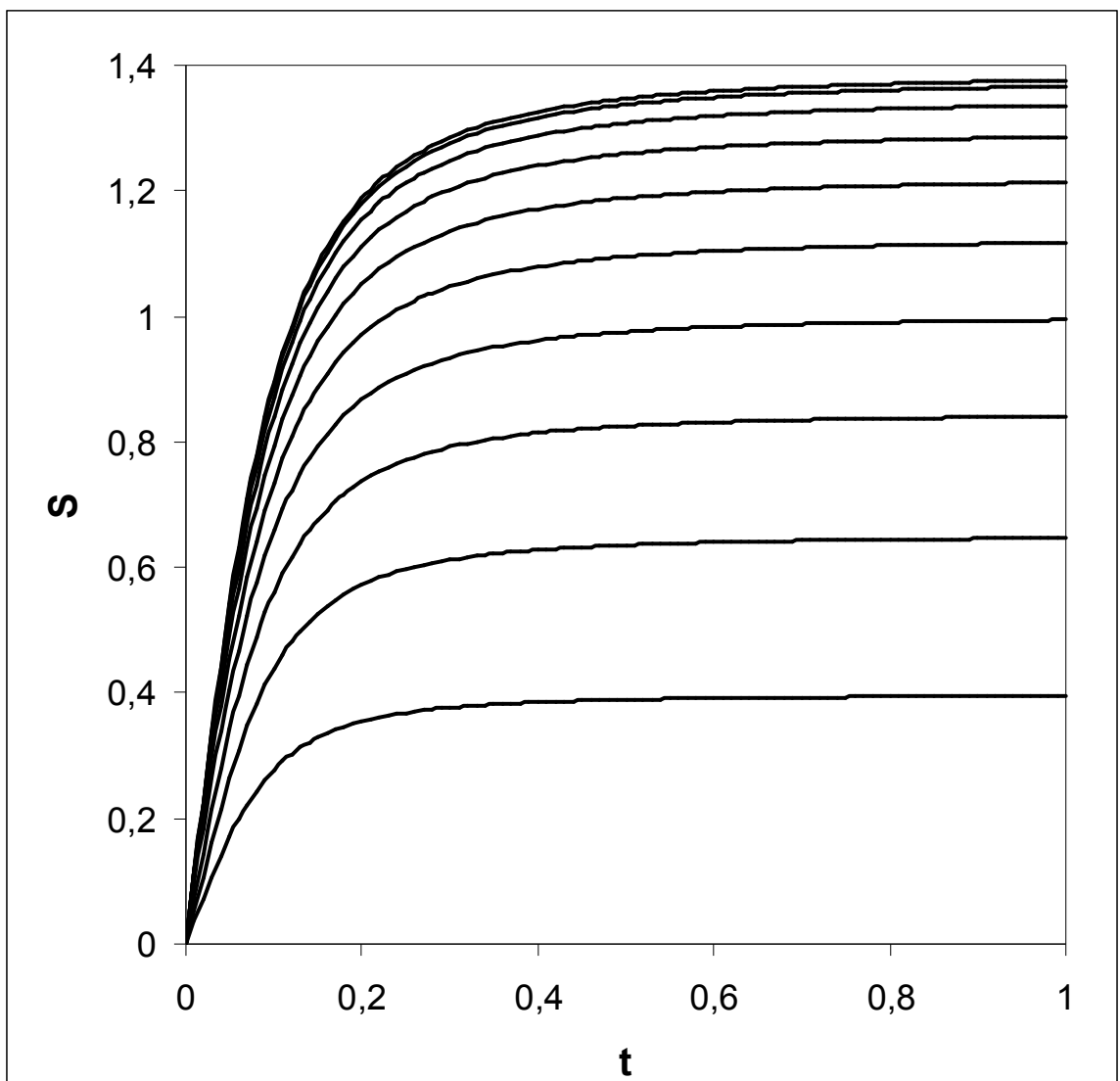

Fig.10
B.V. Karpenko[a], A.V. Kuznetzov[b] and V.V. Dyakin[a]
The tight binding approximation and thermodynamic functions

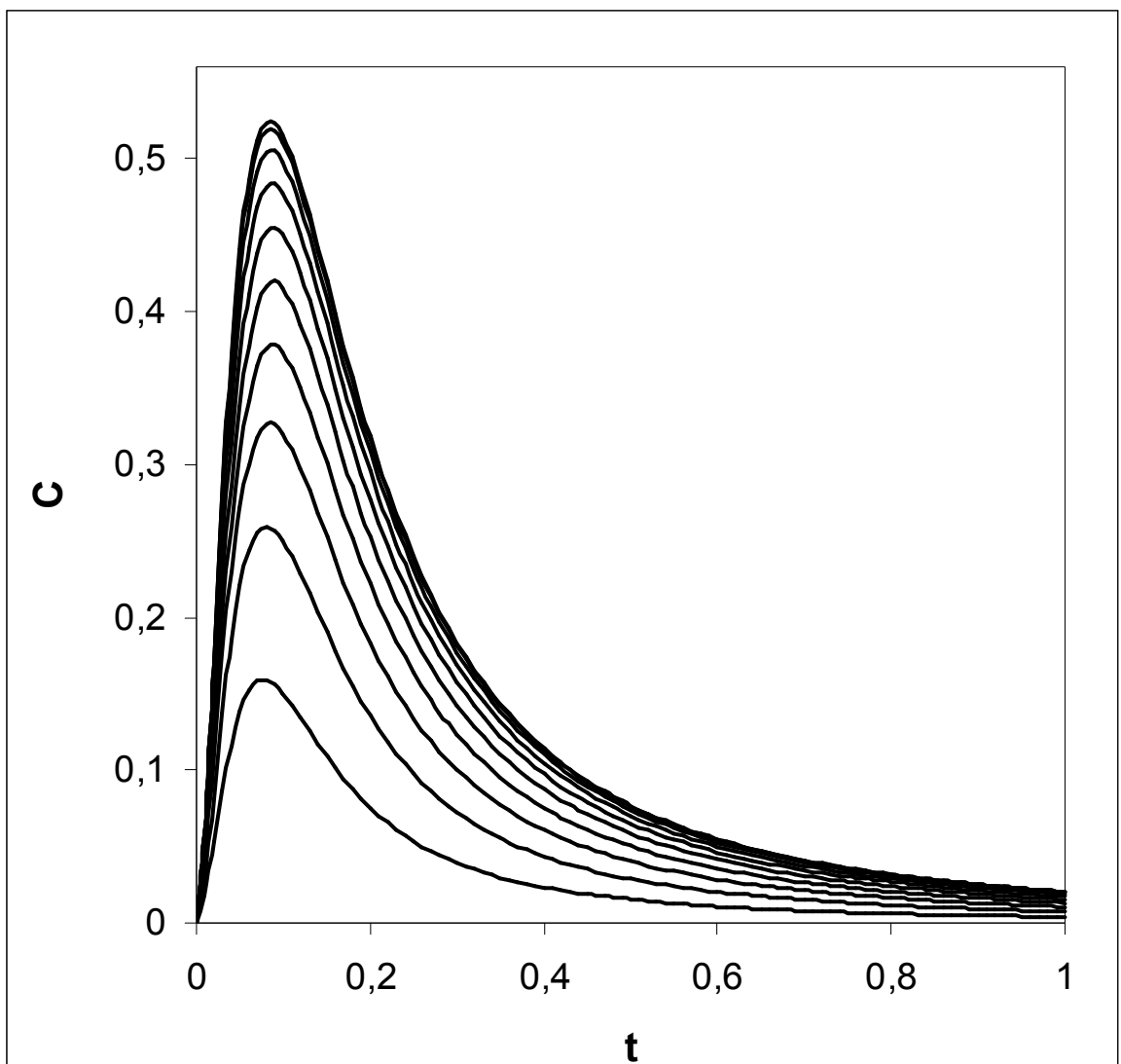

Fig.11
B.V. Karpenko[a], A.V. Kuznetzov[b] and V.V. Dyakin[a]
The tight binding approximation and thermodynamic functions

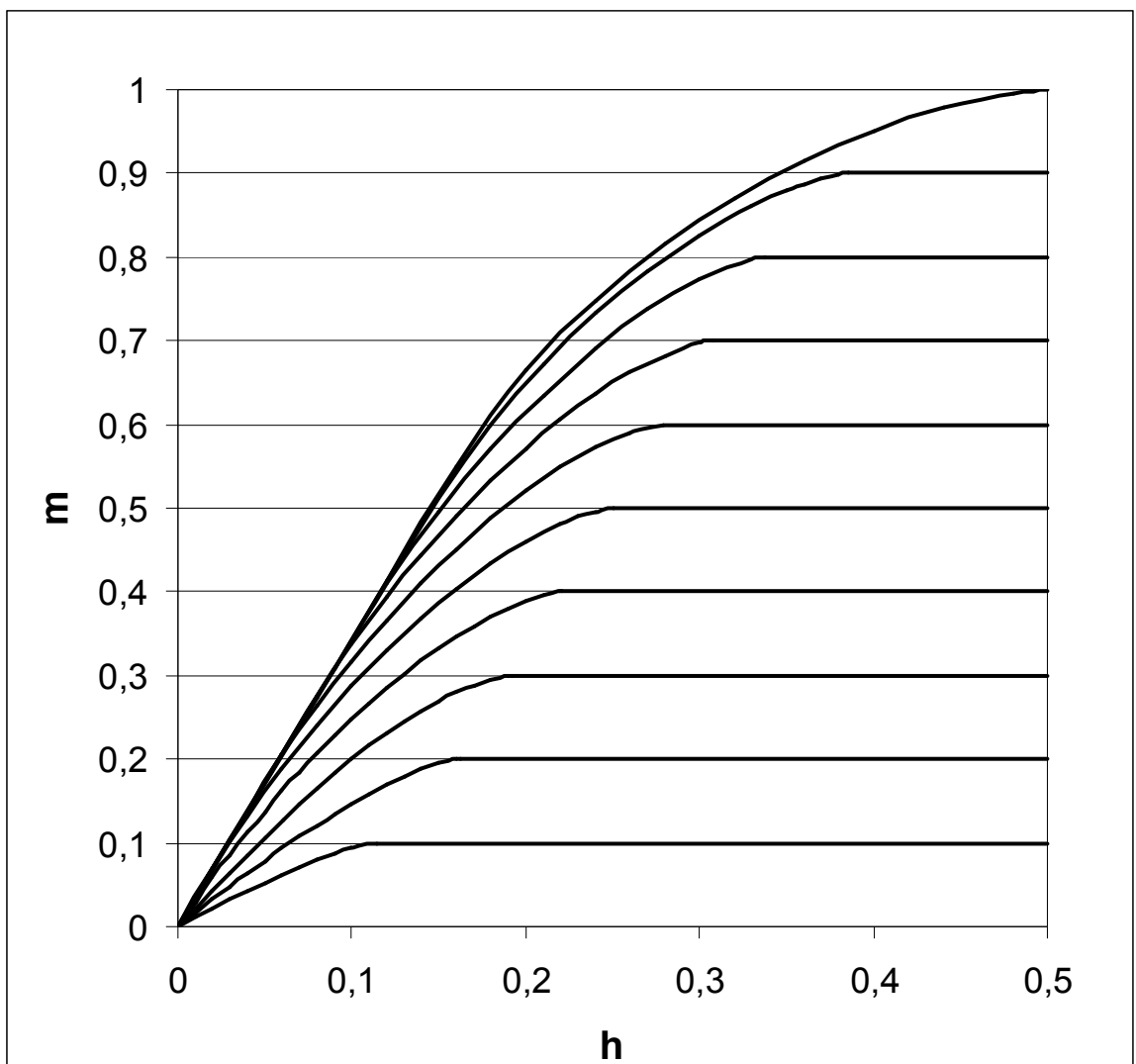

Fig.12
B.V. Karpenko[a], A.V. Kuznetzov[b] and V.V. Dyakin[a]
The tight binding approximation and thermodynamic functions

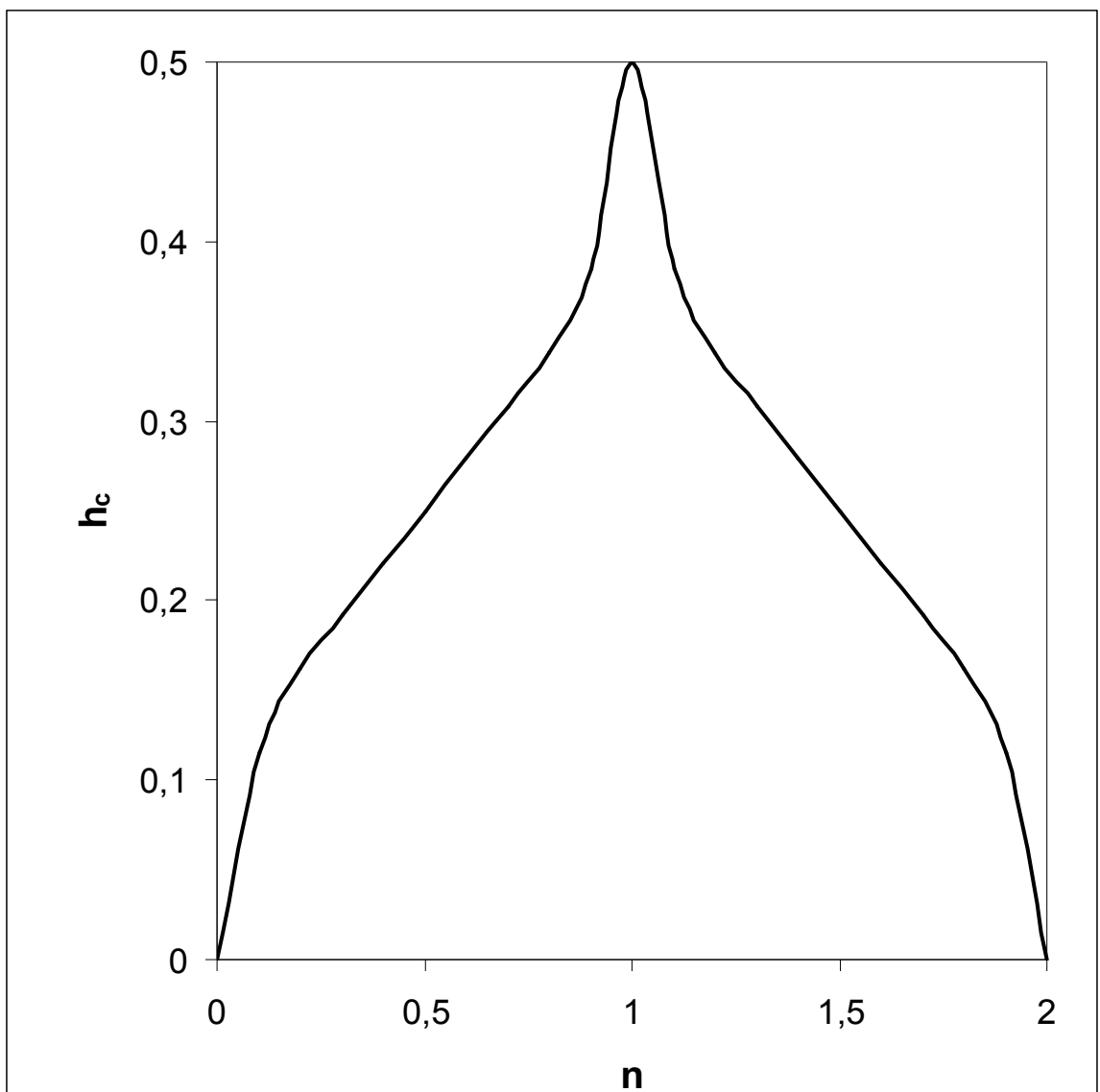

Fig.13
B.V. Karpenko[a], A.V. Kuznetzov[b] and V.V. Dyakin[a]
The tight binding approximation and thermodynamic functions

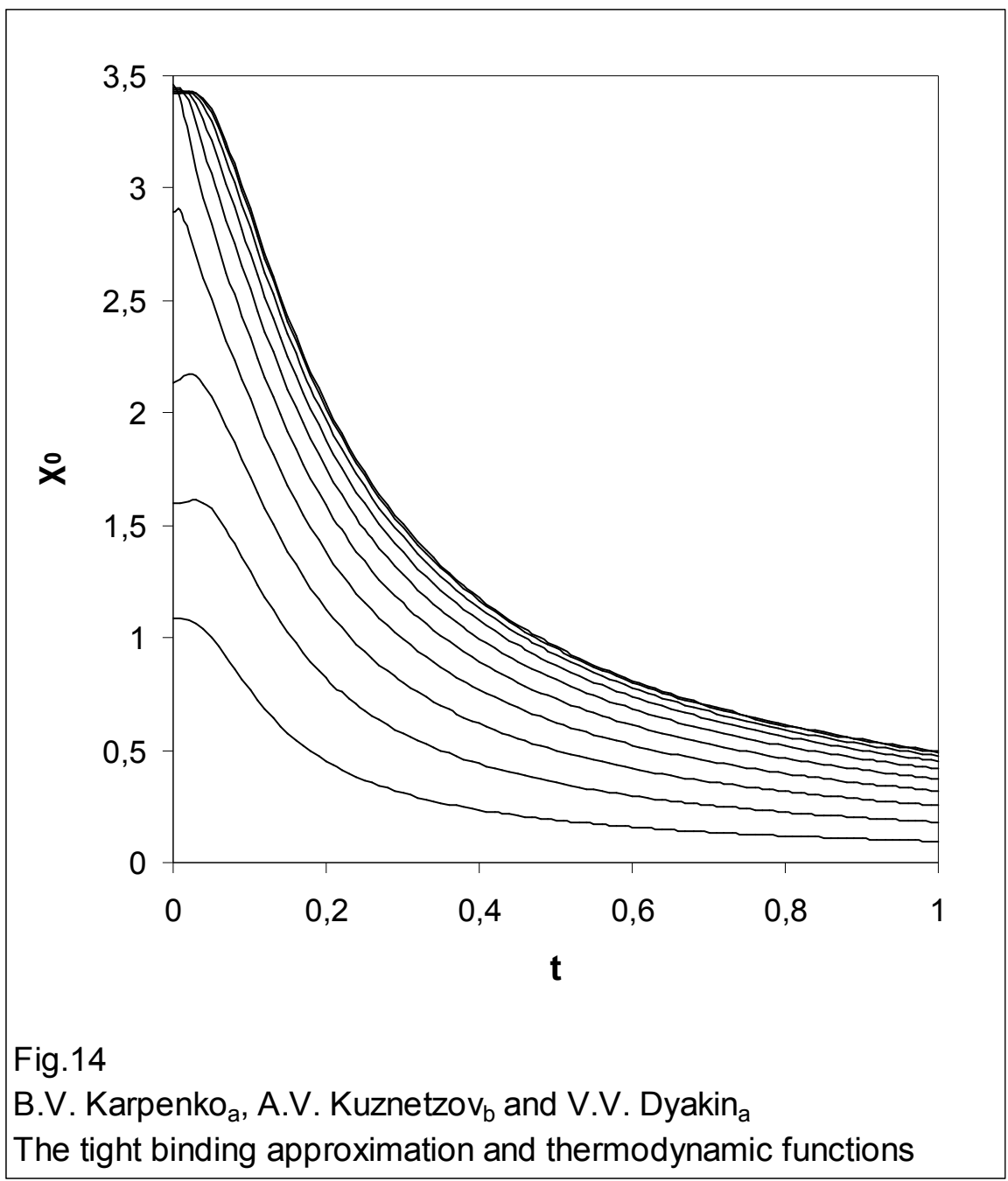

Fig.14
B.V. Karpenko[a], A.V. Kuznetzov[b] and V.V. Dyakin[a]
The tight binding approximation and thermodynamic functions

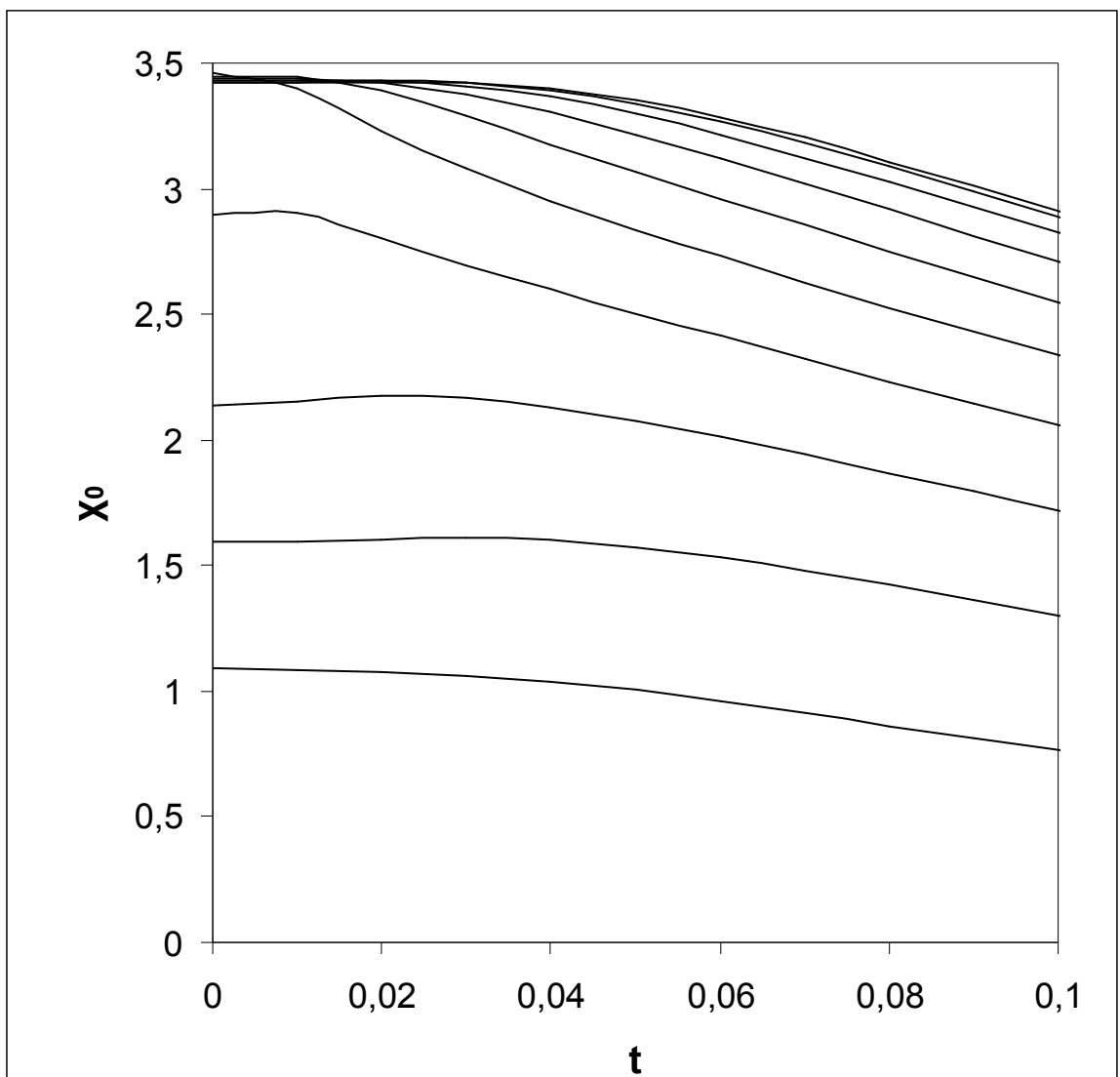

Fig.15
B.V. Karpenko[a], A.V. Kuznetzov[b] and V.V. Dyakin[a]
The tight binding approximation and thermodynamic functions

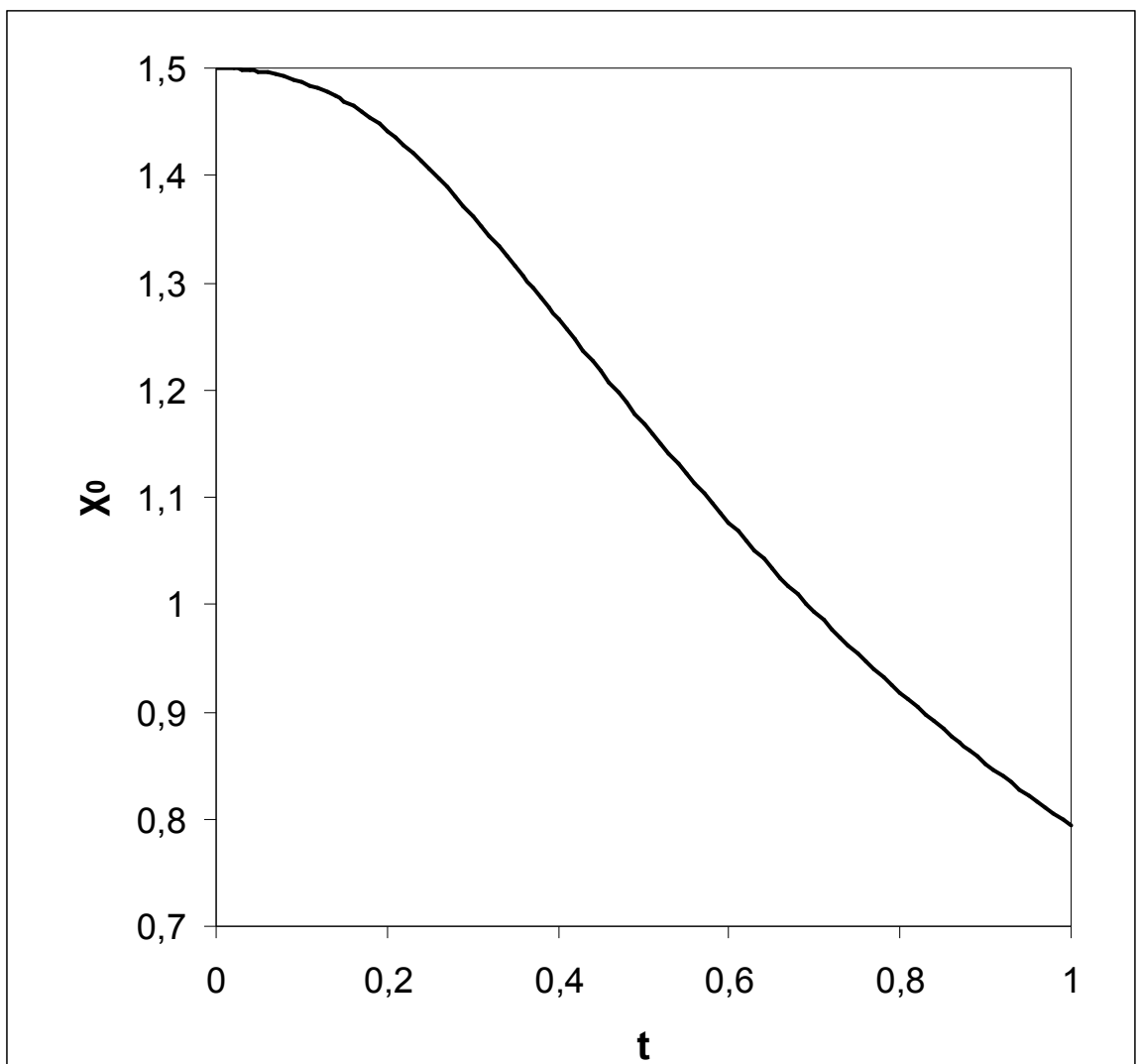

Fig.16
B.V. Karpenko[a], A.V. Kuznetzov[b] and V.V. Dyakin[a]
The tight binding approximation and thermodynamic functions